\documentclass{aa}  

\usepackage{graphicx}
\usepackage{txfonts}
\usepackage{xcolor}
\usepackage[colorlinks=true,linkcolor=blue,citecolor=blue]{hyperref}
\usepackage{gensymb}
\usepackage{subfigure}
\usepackage{amsmath}
\usepackage{orcidlink}
\usepackage{comment}
\usepackage{subfigure}
\newcommand{\target}[0]{TOI-519 b}
\newcommand{\kcl}[0]{KCl~}
\newcommand{\NaS}[0]{Na$_{\mathrm{2}}$S~}
\newcommand{\mynotes}[1]{\textcolor{red}{#1}}
\begin{document}

   %\title{JWST's observations of warm Jupiters around M-dwarf stars might be the key for understanding clouds and hazes in warm atmospheres}
   %\title{JWST's observations of warm Jupiters around M-dwarf stars are key for extensive chemical, cloud and haze characterisation.}
   \title{Warm Jupiters around M-dwarfs are great opportunities for extensive chemical, cloud and haze characterisation with JWST}

   \author{L. Teinturier\inst{1,2}\thanks{lucas.teinturier@obspm.fr}
          \and
          E. Ducrot\inst{1,3,4}
          \and 
          B. Charnay\inst{1}
          }

   \institute{LESIA, Observatoire de Paris, Université PSL, Sorbonne Université, Université Paris Cité, CNRS, 5 place Jules Janssen, 92195 Meudon, France
         \and
            Laboratoire de Météorologie Dynamique, IPSL, CNRS, Sorbonne Université, Ecole Normale Supérieure, Université PSL, Ecole Polytechnique, Institut Polytechnique de Paris, 75005 Paris, France 
            \and 
            AIM, CEA, CNRS, Université Paris-Saclay, Université Paris Cité 91191 Gif-sur-Yvette, France 
            \and 
            Astrobiology Research Unit, Universit\'e de Li\`ege,  All\'ee du 6 ao\^ut 19, Li\`ege, 4000, Belgium
             }

   \date{Received XXX, ; accepted XXX}

% \abstract{}{}{}{}{} 
% 5 {} token are mandatory
 
  \abstract
   {The population of short-period giant exoplanets around M-dwarf stars is slowly rising. These planets present an extraordinary opportunity for atmospheric characterisation and defy our current understanding of planetary formation. Furthermore, clouds and hazes are ubiquitous in warm exoplanets but their behaviour is still poorly understood.}
   {We study the case of a standard warm Jupiter around a M-dwarf star to show the opportunity of this exoplanet population for atmospheric characterisation. We aim to derive the cloud, haze, and chemical budget of such planets using JWST.}
   {We leverage a 3D Global Climate Model, the \texttt{generic PCM}, to simulate the cloudy and cloud-free atmosphere of warm Jupiters around a M-dwarf. We then post-process our simulations to produce spectral phase curves and transit spectra as would be seen with JWST.}
   {We show that using the amplitude and offset of the spectral phase curves, we can directly infer the presence of clouds and hazes in the atmosphere of such giant planets. Chemical characterisation of multiple species is possible with an unprecedented signal-to-noise ratio, using the transit spectrum in one single visit. In such atmospheres, NH$_{\mathrm{3}}$ could be detected for the first time in a giant exoplanet. We make the case that these planets are key to understanding the cloud and haze budget in warm giants. Finally, such planets are targets of great interest for Ariel.}
   {}

    \titlerunning{The warm giant planets and M-dwarf stars opportunity}
   \maketitle
%
%________________________________________________________________
 
\section{Introduction}
M-dwarfs stars are promising hosts for exoplanets and their atmospheric characterisation, as they are the coolest, smallest, and most abundant stars in our neighbourhood \citep{bochanski_luminosity_2010}. This is referred to as the \textit{small star opportunity} \citep{madhusudhan_exofrontiers_2021}. Thus, a few surveys have been designed to study such systems (such as SPECULOOS \citep{delrez_speculoos_2018,triaud_m_2023}, PINES \citep{tamburo_perkins_2022}, EDEN \citep{gibbs_eden_2020}, MEarth \citep{nutzman_design_2008} or using \texttt{TESS} \citep{ricker_transiting_2015}), leading to major discoveries of rocky planets (for instance the TRAPPIST-1 system, \citep{gillon_temperate_2016,gillon_seven_2017}). While close-in planets orbiting around M stars are rare, at least a dozen close-in giant planets have been discovered \citep{triaud_m_2023}. Due to the favourable ratio of the size of the planet over the size of the star, atmospheric characterisation of such targets is more favourable than around F, G, and K stars. Nevertheless, the question of how these giant planets formed around low-mass stars is still a mystery. Indeed, core accretion models predict a low probability of formation of giant planets around such stars \citep{schlecker_rv-detected_2022}. Thus, other formation pathways, such as gravitational instabilities followed by inward migration, might be at play in these systems \citep{morales_giant_2019,liu_tale_2020,kennedy_planet_2008}. \par

Clouds and hazes are ubiquitous in the atmosphere of giant planets. In particular, clouds (i.e., particles formed through condensation of gases) have multiple absorption signatures in the infrared \citep{grant_jwst-tst_2023, morley_water_2014,dyrek_so2_2024,miles_jwst_2023}  and they're also known to weaken emergent spectral lines \citep{sing_continuum_2016}. Hazes (i.e., particles formed through complex photochemical reactions originating from the stellar UV flux), could explain super-Rayleigh slopes seen in transit spectra \citep{ohno_super-rayleigh_2020} and are thought to dominate over clouds in cooler atmospheres \citep{gao_aerosols_2021,lavvas_aerosol_2017,arfaux_large_2022}. Where clouds produce a greenhouse effect in hot atmospheres \citep{teinturier_radiative_2024}, hazes are thought to produce an anti-greenhouse effect \citep{mckay_thermal_1989}. Using a 3D Global Climate Model (GCM), \cite{steinrueck_photochemical_2023} modelled the radiative feedback of hazes on a Hot Jupiter and found that regardless of the assumed radiative properties of the hazes, a strong dayside temperature inversion forms at low pressure. However, they were unable to match the transmission spectra. For sub-Neptune, featureless transit spectra in the near-infrared, probably due to high thick clouds or hazes have been observed \citep{kreidberg_clouds_2014} and modelled \citep{charnay_3d_2015-1}, but the spectral characterisation of the composition and size distribution of the aerosols are still lacking. Thus, clouds and hazes are still fairly unknown on giant exoplanets. \par 

Now that we entered the JWST era, precise atmospheric characterisation is enabled on Hot Jupiters both in emission and transmission spectroscopy, yielding unprecedented signal-to-noise ratio (S/N) and various atmospheric features \citep{alderson_early_2023,bell_nightside_2024,feinstein_early_2023,rustamkulov_early_2023,jwst_transiting_exoplanet_community_early_release_science_team_identification_2023,tsai_photochemically_2023,coulombe_broadband_2023,mikal-evans_jwst_2023,grant_jwst-tst_2023}. Transit spectroscopy on giant planets around M-dwarf stars is currently undergoing for eight systems (Program 3171, \cite{kanodia_red_2023}, Program 5799, \cite{garcia_toi-3884_2024}, Program 5863, \cite{murray_shining_2024}) but to date, emission spectroscopy and phase curves are not yet scheduled. However, short-period giant planets around M-dwarfs are prime targets for a complete energetic budget, albedo, chemical, clouds, and hazes characterisation using either NIRSPec-PRISM (to probe both thermal and reflected light without saturation of the detector) or in thermal emission with MIRI-LRS. In particular, the M-dwarfs do not saturate the NIRSPec-PRISM detector, which makes them advantageous compared to many exoplanet's host stars for which the PRISM cannot be used. Moreover, the launch of PLATO and Ariel in the coming years will yield longer phase curves than the one obtained by JWST, at optical and near to mid infrared wavelengths.\par 
In particular, as M-dwarf stars are cool, giant planets orbiting them are in the temperate temperature regime \citep{madhusudhan_exofrontiers_2021}. Thus, they can be used as a proxy for hot sub-Neptune with better S/N, Transmission, and Emission Spectroscopic Metrics (TSM and ESM) \citep{kempton_framework_2018}. Therefore, insights into clouds and hazes formation and composition that are hardly constrainable on smaller planets (such as GJ-1214b for instance, \cite{kempton_reflective_2023}), could be obtained with these targets. In this letter, we investigate the potential of atmospheric characterisation with JWST for giant planets around M-dwarf stars using a Global Climate Model. \par
In Section \ref{section: methods} we describe the choice of the target and the numerical models used. We present our results in Section \ref{section: results} and discuss and conclude this work in Section \ref{section: discussion}.

%__________________________________________________________________

\section{Methods}\label{section: methods}

\subsection{Target selection}
\begin{figure}
   \resizebox{\hsize}{!}
            {\includegraphics[width=\textwidth]{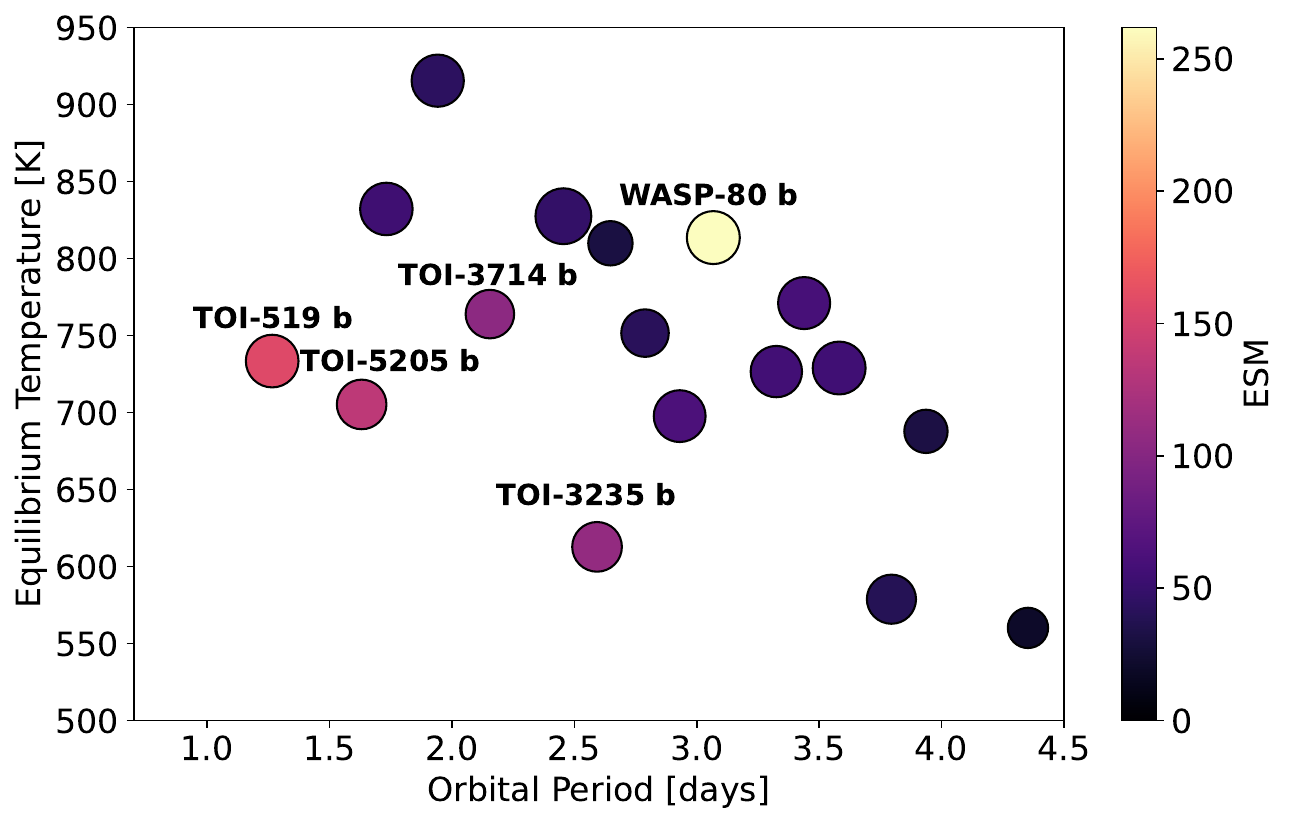}}
      \caption{Population of confirmed close-in gas giants around M-dwarf stars, in the orbital period-equilibrium temperature space. The colourbar represents the ESM and the size of each dot is proportional to the planetary radius. We highlighted the five targets with the best ESM.}
         \label{fig: esm}
\end{figure}
In this letter, we focus on planets around M-dwarf stars with an equilibrium temperature below 800 K. These planets are indeed a good analogue for warm sub-Neptune and Neptune (i.e., GJ 1214 b, GJ 436 b, GJ 3470 b), but with better potential for atmospheric characterisation, due to their size. In this temperature regime, we expect to be able to investigate hazes and clouds simultaneously. We show in Fig. \ref{fig: esm} the population of confirmed (to-date) warm Jupiters around M-dwarf stars, in the orbital period-equilibrium temperature space, colour-coded using the Emission Spectroscopic Metric (ESM, \cite{kempton_framework_2018}). For this study to stay generic, we chose a target with a good enough ESM and an orbital period of less than two days. We use this criterion as observing a phase curve of more than two days seems improbable in the near future with current observing facilities. This leaves us with few choices, even though confirmation of numerous planetary candidates satisfying these criteria is currently ongoing (The SPECULOOS collaboration, private comm., \cite{kanodia_searching_2024}), which will widen the parameter space. Thus, we choose to model \target~\citep{parviainen_toi-519_2021,kagetani_mass_2023,hartman_toi_2023}, as a benchmark of warm Jupiters around M-dwarf stars, amenable for clouds and hazes characterisation. Moreover, we state that \target~is observable with JWST with all instrument modes, without saturation. 

\subsection{General Circulation Model}
\begin{figure*}
   \resizebox{\hsize}{!}
            {\includegraphics[width=\textwidth]{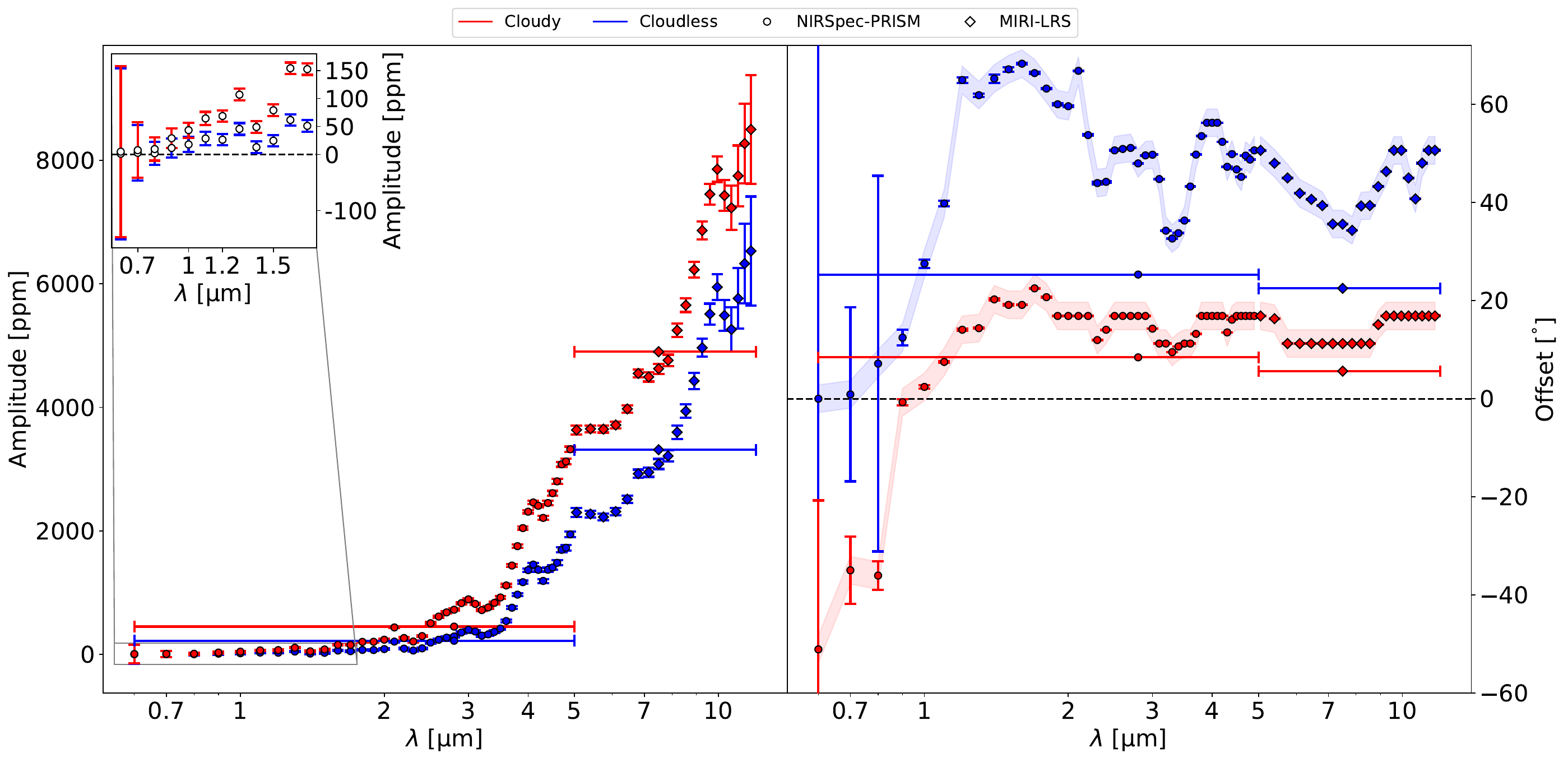}}
    \resizebox{\hsize}{!}
            {\includegraphics[width=\textwidth]{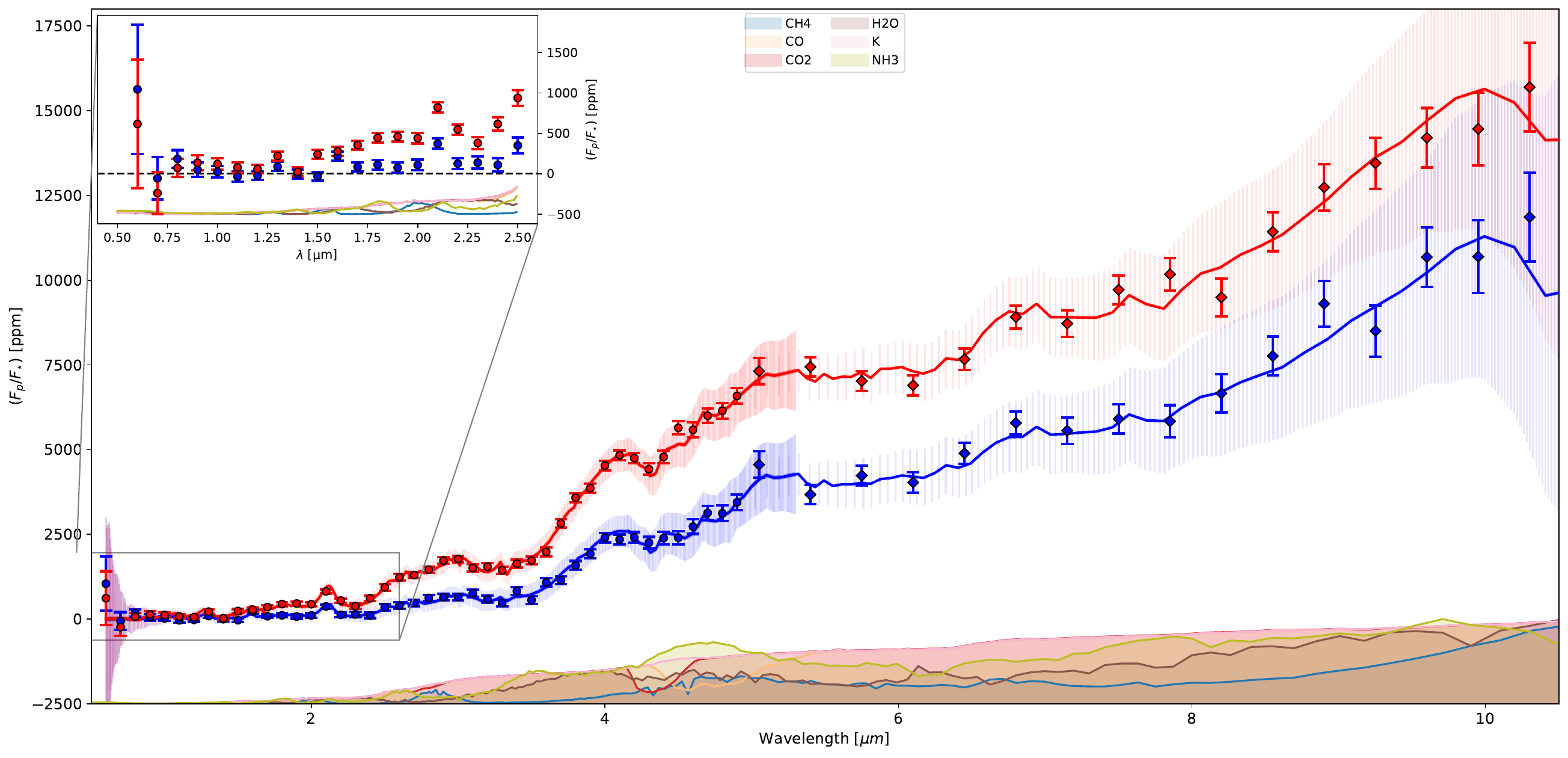}}
      \caption{Spectral features of the phase curve and secondary eclipses.\emph{Top: }Amplitude and offset of the phase curves simulated with NIRSpec-PRISM and MIRI-LRS. \emph{Bottom: } Secondary eclipses on the NIRSpec-Prism and MIRI-LRS range. The dots represent the NIRSpec-PRISM range while the diamonds are the MIRI-LRS range. Cloudless simulations are in blue and cloudy simulations are in red. In all observing modes, the effect of clouds is unambiguously detectable. The zoomed-in subplot allows better visualisation of NIRSpec-PRISM's shorter wavelengths. The lines in which horizontal errorbars encompass the whole instrument bands correspond respectively to the white amplitude and offset of the instrument. Contributions from major chemical species are shown at the bottom of the plot and of the zoomed-in subplot. On the right panel, the shaded region corresponds to a constant error (introduced by the model's coarse horizontal resolution) of $\sim$2.8125$^{\degree}$ on the offset location, as each longitudinal grid cell spans an angular resolution of 5.625$^{\degree}$. On the bottom panel, the shaded regions around the spectra correspond to the native resolution of the instruments and their errors.}
         \label{fig: ampoff}
\end{figure*}

\subsection{Simulation set-up}

We use the \texttt{generic PCM} to model \target, around its M3.5 star (T$_{\star}$ = 3300 K, R$_{\star}$ = 0.35 R$_{\odot}$, \cite{kagetani_mass_2023}). The planet is modelled on a tidally locked, circular orbit at a distance of 0.015 AU, with an orbital period of 1.265 days \citep{kagetani_mass_2023}. The planetary radius R$_{\mathrm{p}}$ and the gravity are fixed at 1.03 R$_{\mathrm{Jup}}$ and 10.76 m.s$^{\mathrm{-2}}$ respectively, according to the values derived in \cite{kagetani_mass_2023}. We also use the value of 2.346 g.mol$^{\mathrm{-1}}$ and 12305 J.K$^{\mathrm{-1}}$.kg$^{\mathrm{-1}}$ for the mean molecular weight and the specific heat capacity of the atmosphere. The internal temperature T$_{\mathrm{int}}$ is set to 100 K. \par
We initialise the model following the procedure described in \cite{teinturier_radiative_2024} and briefly summarised hereafter. We use the \texttt{Exo-REM} code \citep{charnay_self-consistent_2018,blain_1d_2021} to compute vertical chemical profiles for H$_{\rm 2}$O, CO, CH$_4$, CO$_2$, FeH, HCN, H$_2$S, TiO, VO, Na, K, PH$_3$ and NH$_3$ assuming disequilibrium chemistry and a solar metallicity, and a vertical temperature profile (see Fig. \ref{fig:exorem}). The disequilibrium processes are computed using an analytical formalism, which compares the chemical time constants with the vertical mixing time, taken from \cite{zahnle_methane_2014}, with an eddy mixing coefficient of 4.5 $\times$10$^7$ m$^2$.s$^{\rm -1}$. We use these profiles and the \texttt{exo\_k} package \citep{leconte_spectral_2021} to create mixed, temperature-pressure dependent, k-tables with 16 Gauss-Legendre quadrature points. These k-tables and the temperature profile are fed to the 1D version of the \texttt{generic PCM} until reaching radiative balance. The outputted temperature profile is then used to initialise the 3D model, in a horizontally uniform way. \par
We use a horizontal resolution of 64x48 and 40 vertical layers, equally spaced in log-pressure between 80 bars and 10 Pa. The dynamical time-step is 28.76 s and the physical-radiative time-step is 143.81 s. Overall, the set-up is highly similar to the one used in \cite{teinturier_radiative_2024}. \par
Starting from a rest state (i.e., no winds), we run a cloudless model for 3,000 planetary years until the upper atmosphere reaches a steady state. We then add clouds composed of \kcl and \NaS with a mean radius of 1 \textmu m and a log-normal distribution in size of variance 0.1, and run the simulation for an additional 500 planetary years. To initialise the clouds, we use the formulas of \cite{morley_neglected_2012} for the condensable vapour, and start with no clouds (see Fig. \ref{fig:cloud_init}). The optical properties used are taken from \cite{querry_optical_1987} for \kcl and from \cite{montaner_optical_1979} and \cite{khachai_fp-apwlo_2009} for \NaS. We only consider these two cloud species as they are the expected clouds to form in this temperature range. We neglect ZnS, which could also form, as the optical properties of ZnS clouds have negligible radiative contributions compared to the other two. \\
The cloud parametrisation used is the same as in \cite{teinturier_radiative_2024}, where at each time-step and in each grid cell, the saturation vapour pressure of the cloud species is computed using the condensation curves of \kcl and \NaS and compared to the partial pressure of the gas. If the grid cell is saturated, condensable vapour condenses into a solid. Clouds can then be transported by the dynamics and/or experience sedimentation. Latent heat release from cloud condensation is also taken into account but is negligible in these atmospheres.

\subsection{Simulations of observations using PandExo}
We post-process our simulations' outputs using the \texttt{Pytmosph3R} code \citep{falco_toward_2022}. We use the same opacities and the same spectral range as for the GCM runs but at a higher spectral resolution equivalent to R $\sim$ 500. Thus for each GCM run, we produce spectral phase curves and a transit spectrum. \\ 
We then use the online version of  \texttt{PandExo} \cite{batalha_pandexo_2017} to estimate the noise of a single transit/eclipse as seen by JWST using NIRISS-SOSS, NIRSpec-PRISM, MIRI-LRS. For the stellar spectrum, we used the default PHOENIX model for T$_{\rm eff}= 3322$ K, normalised to a $J$-mag = 12.847. Table \ref{tab:Pandexo} summarises the parameters that we used to compute the noise. For each instrument, we ran simulations using the cloudless and cloudy transmission and emission spectra output from the GCM runs. 
\begin{table}[]
    \centering
    \begin{tabular}{|p{1.7cm}|p{1.7cm}|p{1.5cm}|p{1.5cm}|}
    \hline
        Instrument &  Groups per integration & Exposure time (s) & Precision white light curve (ppm/min)\\
        \hline
        NIRISS-SOSS & 84 & 466 & 34 \\
        \hline
        NIRSpec-PRISM & 6 & 1.6 & 190 \\
        \hline
        MIRI-LRS & 469 & 74.7 & 1272\\
    \hline
    \end{tabular}
    \caption{Parameters used to compute the expected spectral precision on the transit/eclipse depth of \target using either NIRISS-SOSS, NIRSpec-PRISM or MIRI-LRS. }
    \label{tab:Pandexo}
\end{table}
We also computed the photometric precision of the white light curve to derive our uncertainty of the timing offset of the phase curve of the planet for each instrument, following the methodology of \cite{alegria_uncertainty_2006}.
\section{Results}\label{section: results}
%-------------------------------------------------
\subsection{Atmospheric structure}
%-------------------------------------------------

\begin{figure*}
    \centering
    %\subfigure[Optical offsets]{\includegraphics[width=0.45\textwidth]{Figures/optical_offset.pdf}}
    %\subfigure[Cloud column]{\includegraphics[width=0.45\textwidth]{Figures/daysideclouds.pdf}}
    \includegraphics[width=\textwidth]{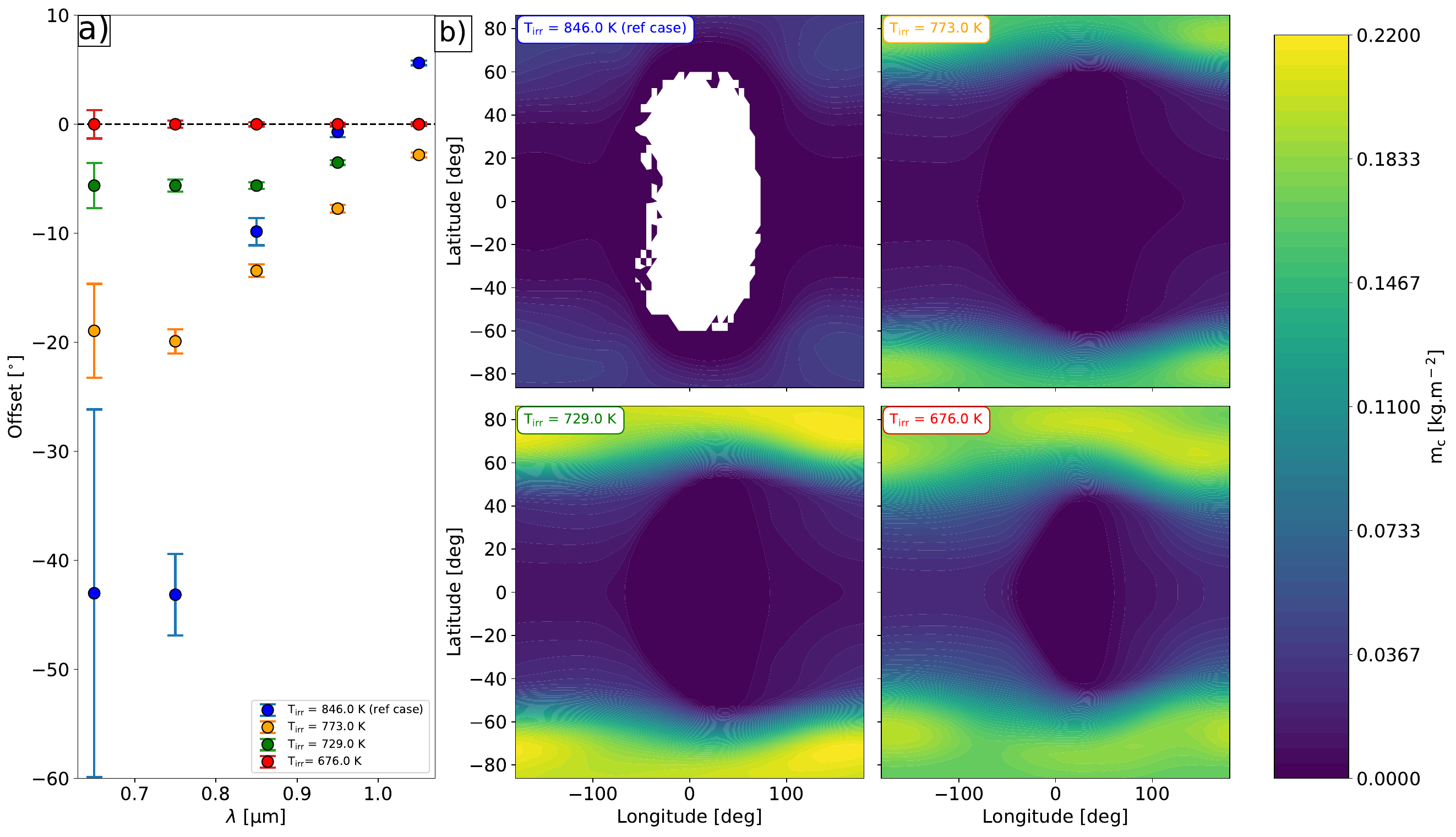}
    \caption{Phase curve offset and link to the dayside cloud structures for different irradiation temperatures. \emph{(a): }NIRSpec optical wavelengths phase curve offsets. Simulations with different irradiation temperatures are represented with different colours. \emph{(b): }Column mass of clouds (both \NaS and \kcl~are considered in these maps), for the same irradiation temperatures as in the left panel.}
    \label{fig:optical}
\end{figure*}

\begin{figure*}
   \resizebox{\hsize}{!}
            {\includegraphics[width=\textwidth]{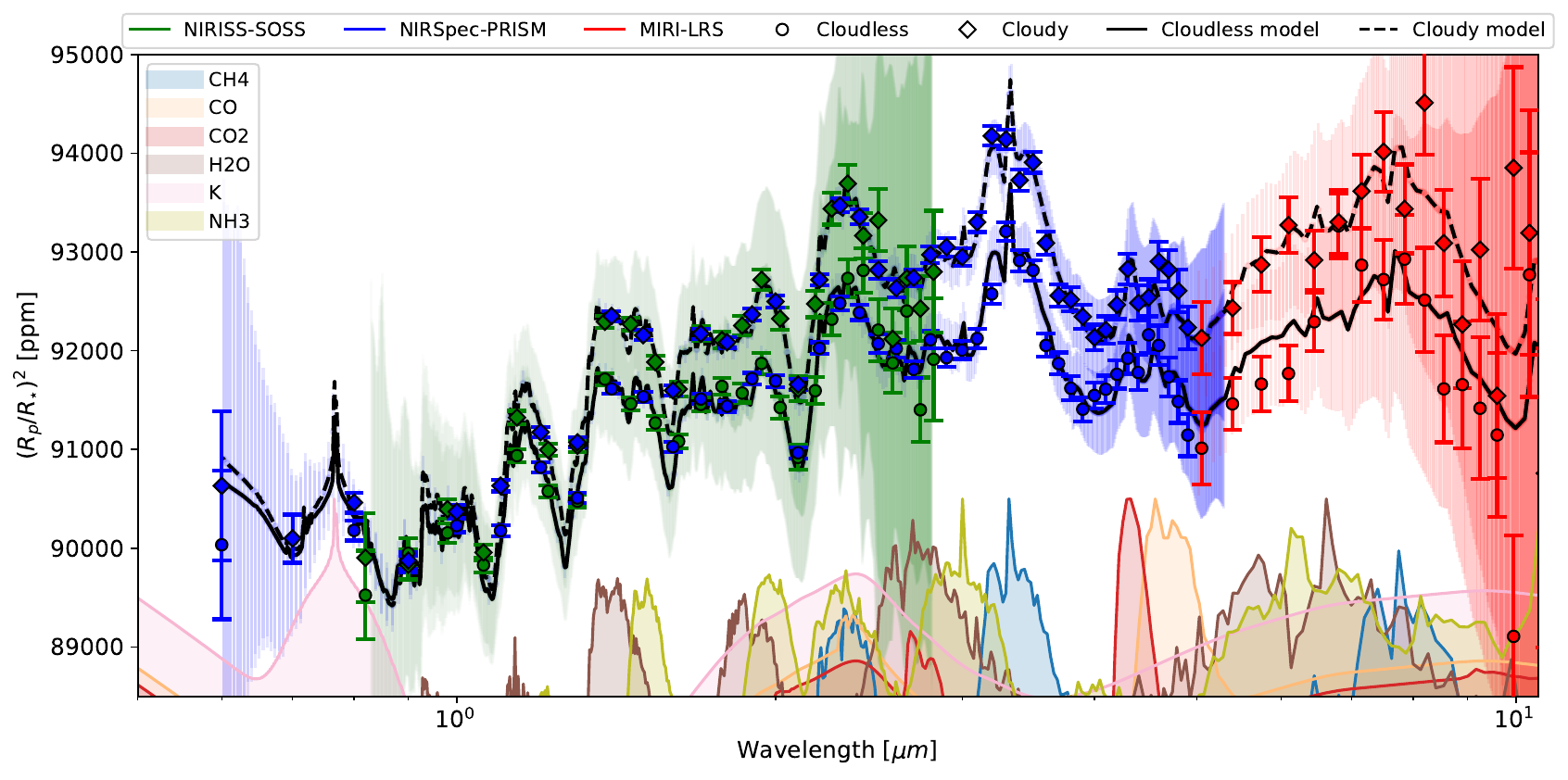}}
      \caption{Transit spectra for our cloudless (dots) and cloudy (diamonds) simulations, noised with \texttt{PandExo}, for NIRISS-SOSS (green, with a bin width of 0.08 \textmu m, NIRSpec-PRISM (blue, with a bin width of 0.1 \textmu m) and MIRI-LRS (red, with a bin width of 0.35 \textmu m) corresponding to one transit with each instrument. The shaded regions around the spectra correspond to the native resolution of the instruments and their errors, before binning. The GCM post-processed transit spectra are in black, with a solid line for the cloudless case and a dashed line for the cloudy case. Contributions from major chemical absorbers are shown at the bottom of the plot.}%\mynotes{We use a random gaussian tirage added to the bin\_y value with sigma = noise}}
         \label{fig: transit}
\end{figure*}
The atmospheric states obtained at the end of the simulations are similar to that of Hot Jupiters around G, F, or K stars (see \cite{showman_atmospheric_2020} for a review). We show in Fig. \ref{fig: thermal_wind_maps} a longitude-latitude temperature map and the structure of the zonal wind for our cloudless and cloudy simulations. In both cases, an equatorial super-rotating jet is present and redistributes heat from the heated dayside to the cool nightside. An eastward shift of the hotspot from the substellar point is also seen, as is the strong temperature contrast between day and nightside. The impact of clouds on the thermal and dynamical structure is very similar to the findings of \cite{teinturier_radiative_2024}, where the clouds have a net warming effect on the atmosphere (nightside warming by greenhouse effects dominate the cooling by albedo effect on the dayside), reduce the eastward shift of the hotpost and decelerate the equatorial jet. Also seen is a modification of the structure of the high-latitude winds, which transition from prograde to retrograde jets, due to the modification of the thermal gradients of the atmosphere by the radiative feedback of the clouds. We show the horizontal distribution of clouds in Fig. \ref{fig: clouds_maps}, on a representative isobaric level. Clouds mostly form on the nightside and at mid-to-high latitudes, leaving a permanent low-latitude cloudless dayside.
%-------------------------------------------------
\subsection{Phase curves and secondary eclipses}
%--------------------------------------------------
We first compare the white light curve amplitude and offset derived from our two simulations for NIRSpec-PRISM and MIRI-LRS. As expected from previous studies \citep[for instance]{parmentier_cloudy_2020}, the phase curve amplitude increases and the offset decreases when clouds are added to the simulations.However, in this particular case of a giant planet around an M-dwarf star, the statistical significance with which these changes can be detected is greater than that for FGK hosts (see Fig. \ref{fig: ampoff}). This change is especially remarkable in the amplitude of the MIRI-LRS phase curve and the offsets with both instruments. Thus, even without leveraging the spectral dependency of the phase curve, such measurements should be sufficient to indicate the presence or lack of clouds. \par

Digging into the spectral shape of the phase curve (which probes different vertical depths), we observe drastic changes for wavelength greater than 2 \textmu m. Indeed, the spectral amplitudes of cloudy simulations are considerably stronger than in the cloudless case, and the spectral offsets subsequently lower. In the MIRI-LRS range, the offset is below 20$^{\degree}$ whereas it is at least twice as large in the clear-sky simulation. This is also the case for NIRSpec-PRISM, with a stronger spectral variability. At shorter wavelengths ($\lambda$ $\leq$ 2 \textmu m) the amplitude level is low ($\leq$ 200 ppm) but is still distinguishable between the two cases presented here. Especially, negative offsets at optical wavelengths are seen, due to the reflected light contribution that is enhanced in the presence of clouds. However, NIRSPec-PRISM's precision at optical wavelength impedes a clear detection of negative offsets as an indicator of cloud scattering contribution. Indeed, the reflected light component is of the order of $\sim$10 ppm, which is the order of magnitude of the thermal component at these wavelengths. For completeness, we show a subset of spectral phase curves converted to brightness temperature in Appendix \ref{appendix: phasecurve}.   \par 
The secondary eclipses are shown for NIRSpec-PRISM and MIRI-LRS on the bottom of Fig. \ref{fig: ampoff}. Multiple spectral features are seen with both instruments. At shorter wavelength, the flux level is clearly detectable above 0.8 \textmu m and is twice as much as the flux observed on WASP-80 b \citep{bell_methane_2023}. For wavelength above 3 \textmu m, the effect of clouds is seen as an increase of flux with regard to the cloudless case. Thus, warm giants around M-dwarf stars are prime target for emission spectroscopy with JWST. 
%-----------------------------------------------------------------
\subsection{Using the optical offset for haze detection}
%---------------------------------------------------------------
As shown in Fig.\ref{fig: ampoff} for \target, negative offsets due to clouds are observed for wavelength below 1 \textmu m, as the dayside is cloudless and the nightside is cloudy (Fig.\ref{fig: clouds_maps}). Thus, if one observes null offsets in the optical range, this would indicate the presence of dayside reflective materials, such as hazes. We investigate the limit of this hypothesis by running additional simulations of a cloudy \target, but lowering the irradiation temperature (thus cooling the planet), until cloud formation at the substellar point. Dayside clouds appear for an irradiation temperature of around 775 K. The geometric albedo increases by 5, 11 and 18 \% for the simulations shown in Fig. \ref{fig:optical} compared with the reference case. However, null optical offsets require more dayside clouds and are seen for lower irradiation temperatures (Fig.\ref{fig:optical}). This behaviour is understandable as when we start to form dayside clouds, we already have an opaque cloud cover at the western terminator which determines the location of the offset. Thus, the phase curve's optical offset could help constrain haze's existence only for irradiation temperatures above $\sim$700 K. This is the case for TOI-3714 b, TOI-5205 b and \target. For lower irradiation temperature, null optical offsets could be due to either hazes, dayside clouds or both without the possibility of distinguishing between the two as for TOI-3235 b. In the unlikely case where cloud formation is prevented, the presence of hazes over a clear sky could be inferred by  measuring the Bond albedo, as done for GJ 1214 b with a MIRI-LRS phase curve \citep{kempton_reflective_2023}.
%----------------------------------------------------------------
\subsection{Transit spectra}
%--------------------------------------------------------------------

We show in Fig. \ref{fig: transit} the transit spectra computed from our cloudless and cloudy simulations and noised with \texttt{PandExo} for NIRISS-SOSS, NIRSpec-PRISM and MIRI-LRS. In all the spectral ranges shown, (0.6 to 10.5 \textmu m), the cloudy and cloudless simulations are offset by 500-1200 ppm. Thus, the absolute flux level of the transit spectrum is indicative of cloudiness. Especially, NIRSpec-PRISM stands out thanks to its combination of a broad wavelength coverage and excellent precision. \par
The $\sim$3.3 \textmu m methane feature is significantly detected with this instrument, allowing strong chemical constraints to be obtained at the terminator of the planet. CH$_{\mathrm{4}}$ has already been detected in the atmosphere of warm Jupiters, both in transit and emission spectroscopy using NIRCAM-F322W2 grism \citep{bell_methane_2023}. However, high spectral resolution was needed to constrain CH$_{\mathrm{4}}$, mostly because of the low flux level of the observed planet. Here, we argue that the low spectral resolution of NIRSpec-PRISM is sufficient because of the strong transit depth ($\sim$9.3 \% for \target~compared to $\sim$3\% for WASP-80b for instance).  \par 
Water features are also seen at shorter wavelengths for NIRSpec-PRISM and in the MIRI-LRS range. Determination of its abundance could help constrain the C/O ratio of the planet. Interestingly, for our assumed chemical composition, hints of CO and CO$_{\mathrm{2}}$ around 4-4.5 \textmu m make their appearance. Moreover, NH$_{\mathrm{3}}$ displays several absorption features at 1.5, 1.9, 3.0, and 6.2 \textmu m. These features might be small and hidden by overlapping absorption bands of H$_{\mathrm{2}}$O or CH$_{\mathrm{4}}$, for instance. However, detection of NH$_{\mathrm{3}}$ with transit spectroscopy has not been done so far (to our knowledge) and our case study of warm Jupiters around M-dwarf stars seems to be a prime target for such investigation. In particular, NH$_{\mathrm{3}}$ would only be detectable if vertical mixing is active and not in the equilibrium chemistry scenario. Thus, transit spectroscopy could provide information about disequilibrium chemistry mechanisms in these atmospheres. Finally, depending on the spectral binning used with NIRSpec-PRISM, a clear detection and characterisation of the potassium at 0.77 \textmu m is feasible. 
\begin{comment}
\begin{figure*}
   \resizebox{\hsize}{!}
            {\includegraphics[width=\textwidth]{Figures/emission_spectrum_simu.pdf}}
      \caption{Secondary eclipse spectra for our cloudless (dots) and cloudy (diamonds) simulations, noised with \texttt{PandExo}, for NIRSpec-PRISM (left panel, with a bin width of 0.1 $\mathrm{\mu}$m) and MIRI-LRS (right panel, with a bin width of 0.35 $\mathrm{\mu}$m). The GCM post-processed transit spectra is in black, with a solid line for the cloudless case and a dashed line for the cloudy case. \mynotes{We use a random gaussian tirage added to the bin\_y value with sigma = noise. MOre binning for prism for clairyt}}
         \label{fig: eclipse}
\end{figure*}
We show the secondary eclipse spectra in Fig.\ref{fig: eclipse}. 
\end{comment}
\section{Discussion and concluding remarks}\label{section: discussion}
In this work, we investigate the spectral observations of a giant planet around a M-dwarf star, as would be seen with JWST. We show that CH$_{\mathrm{4}}$ and H$_{\mathrm{2}}$O should be easily detected with NIRSpec-PRISM and MIRI-LRS in one single transit. Hints of NH$_{\mathrm{3}}$, CO and CO$_{\mathrm{2}}$ are also detectable. These measurements could help constrain the C/O ratio and help explain the formation pathway of such objects \citep{oberg_effects_2011}. Depending on the value of the C/O ratio and C/H and O/H ratio, constraints on the migration mechanism could also be put \citep{madhusudhan_toward_2014}. \par
Measuring a spectral phase curve will allow unprecedented constraints on the cloud and chemical budget of these giant planets. The spectral amplitude and offset will reveal the presence/absence of clouds. Depending on the target, one observing mode might be more favourable than the other. For the special case of \target, we first recommend a MIRI-LRS phase curve followed by a NIRSpec-PRISM one, as the S/N of these observations is high. Furthermore, it is worth noting that if the LRS fixed-slit mode demonstrates significantly higher sensitivity (as proposed by the calibration program GO 6219, PIs Dyrek \& Lagage, \cite{dyrek_miri_2024}) compared to the slitless mode for cooler and fainter stars, such as TOI-519, we may anticipate even greater precision. This enhanced precision would enable us to discern an even wider range of atmospheric scenarios.  \par
In this work, we did not consider the G395H filter despite its wide use for atmospheric characterisation \citep{alderson_early_2023,smith_combined_2024}. As shown previously, the lower spectral resolution of the PRISM is sufficient enough to yield a clear detection of CH$_{\mathrm{4}}$ and also allows for probing a large spectral range including reflected light. Thus, using the G395H filter would yield great constraints on the carbon chemistry budget in these atmospheres but the PRISM is more adequate on the first hand because of its broad wavelength coverage. \par
Finally, we do not consider hazes in our modelling, which should be present on the dayside of such planets due to photochemistry \citep{steinrueck_photochemical_2023}. As shown in Fig. \ref{fig: clouds_maps}, our simulated dayside is cloudless. Thus, if observations yield a null offset of the phase curve at optical wavelength, this is a strong evidence of the presence of such hazes, whereas negative optical offsets will inform us on a clear (of clouds and hazes) dayside and a cloudy western terminator. However, our additional simulations highlight the existence of a thermal transition from negative to null optical offset due to an enhanced abundance of dayside clouds.\par
We also didn't consider stellar activity which could mitigate the goodness of these observations, as M-dwarf stars are known to exhibit spots and flares \citep{rackham_effect_2023}. Thus, our estimated spectra are optimistic. \par
We emphasise that the climate model used in this study, the \texttt{generic PCM} is well suited for flexible and fast simulations of giant planets around any stellar type. In this letter, we present the first 3D atmospheric study of a giant planet around a M-dwarf star. \par
Finally, the growing population of giant planets around M-dwarf stars are prime candidates for the Ariel space telescope \citep{tinetti_chemical_2018}, which will survey more than 1000 exoplanets and dedicate 10\% of its time to phase curves \citep{charnay_survey_2021}. Ariel's spectral coverage from 0.7 to 7.8 \textmu m combined with its possibility to observe relatively long phase curve (up to 5 days), make it ideal for probing this population of exoplanets.
\begin{comment}
\begin{itemize}
    \item hazes not taken into account 
    \item stellar activity as well, but should be okay on a M3
    \item CH4 should be detected easily => chemical budget should be interesting to constrain =>> wasp-80 b paper, and H2O as well => C/O and formation constraints
    \item One phase curve with JWST will allow unprecedented constraints on the cloud budget of these Hot Jupiters, via the transit spectrum and the white and spectral phase curves. Depending on the chosen targets, one observing mode might be favourable, but for \target, we recommend a \mynotes{je pense MIRI pour l'amplitude enorme de la phase curve} as the SNR of such observations is amongst the strongest exoplanetary signal, allowing a good atmospheric characterisation \mynotes{cette phrase veut rien dire mais les idées sont là}
    \item parler du G395H. Si on a une bonne caracterisation avec prism et sa resolution pourrie, on l'aura d'autant plus avec celui-là. Mais on perd la lumière réfléchie
    \item We also emphasise that our model allow for flexible and fast simulations of any targets. Firs 3D study of Hot J around M dwarfs
    \item ARIEL -> prime targets
\end{itemize}
\end{comment}
\begin{acknowledgements}
This work has been granted access to the HPC resources of MesoPSL financed byb the Region Ile de France and the project Equip@Meso (reference ANR-10-EQPX-29-01) of the programme Investissements d’Avenir supervised by the Agence Nationale pour la Recherche. This work was supported by CNES, focused on AIRS on Ariel. E. D. acknowledges support from the innovation and research Horizon 2020 program in the context of the Marie Sklodowska-Curie subvention 945298. L.T and E.D strongly thank the creators of \url{https://makeitmeme.com} for bringing a lot of joy into the writing of this paper. The authors thank the anonymous referee for constructive remarks that enhanced the quality of this work.
\end{acknowledgements}

%-------------------------------------------------------------------

\bibliography{references}
\bibliographystyle{aa}
\begin{appendix}

\section{Initialisation}\label{appendix: init}
\begin{figure}[h]
    \centering
    {\includegraphics[width=0.5\textwidth]{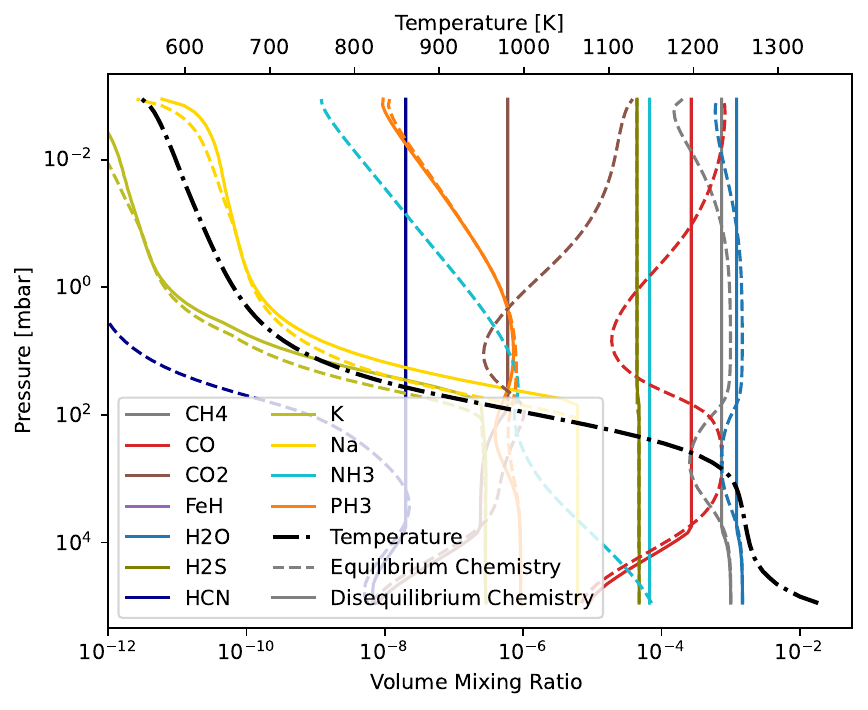}}
    \caption{Chemical (coloured lines) and temperature (black dotted line) profiles derived from \texttt{Exo-REM}. The solid lines correspond to disequilibrium abundances and the dashed lines to chemical equilibrium abundances. The atmosphere is dominated by CH$_4$ and not CO because of the low temperature. Although they are included in the computation, TiO and VO mixing ratios are negligible and do not appear on the plotted scale. }
    \label{fig:exorem}
\end{figure}

\begin{figure}[h]
    \centering
    {\includegraphics[width=0.5\textwidth]{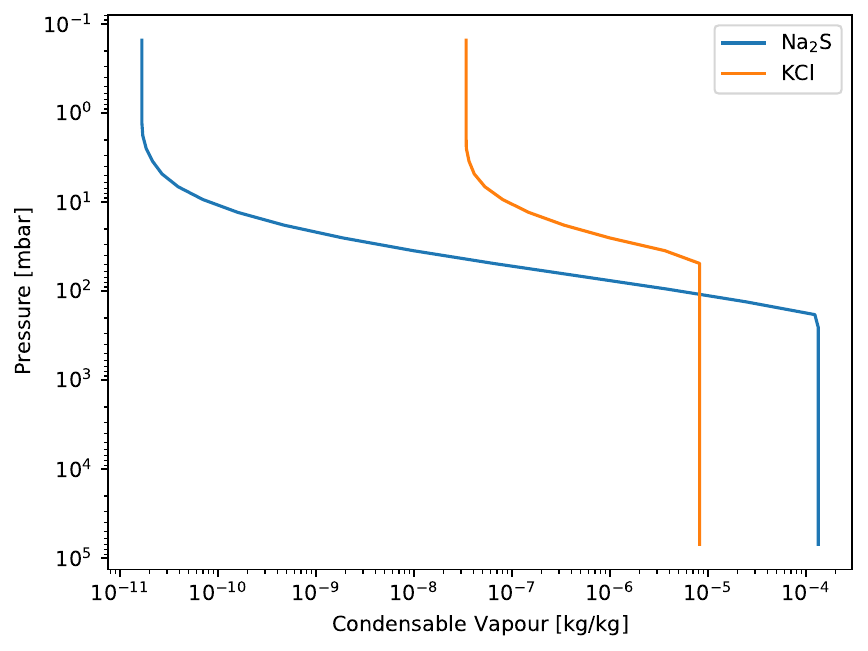}}
    \caption{Condensable vapour profile of \kcl and \NaS used at initialisation. We start with no clouds and only condensable vapour in the atmosphere.}
    \label{fig:cloud_init}
\end{figure}
\section{Atmospheric states}
\begin{figure*}
   \resizebox{\hsize}{!}
            {\includegraphics[width=\textwidth]{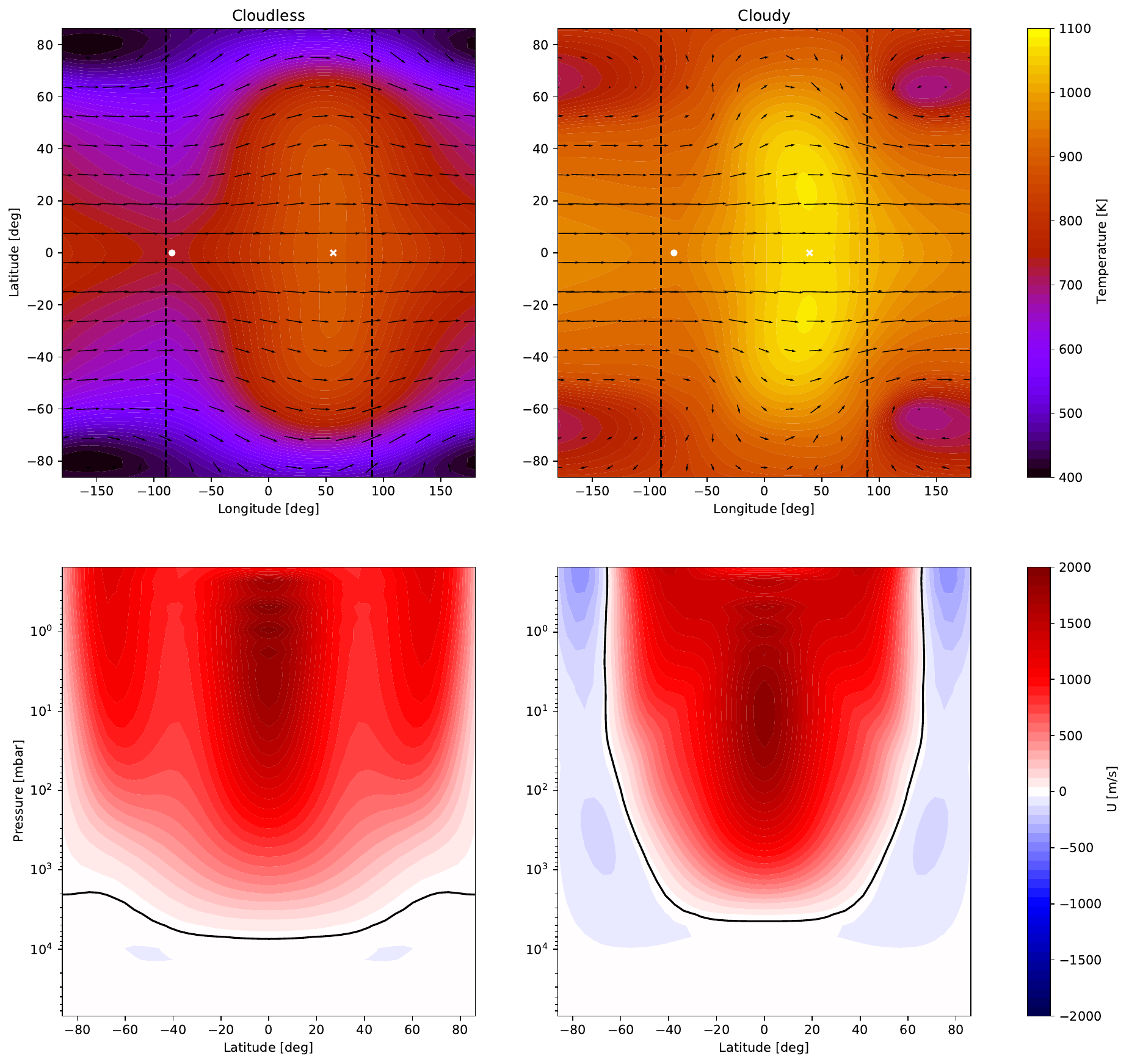}}
      \caption{Thermal and wind structure of our cloudless and cloudy simulations of \target. \emph{Top: }longitude-latitude temperature map at a pressure of 25 mbar, averaged in time over the last 100 years. The black arrows represent the wind field, with the size of the arrow being proportional to the strength of the wind. The vertical dashed black lines denote the terminators (longitude = $\pm$90$^{\degree}$). The white cross and dot represent the equatorial maxima and minima of temperature. \emph{Bottom: } Zonal mean zonal wind (averaged in time over the last 100 years). The black line denotes the 0 m.s$^{\mathrm{-1}}$ level. The left panels represent the cloudless simulation while the right panels represent the simulation including \kcl and \NaS clouds.}
         \label{fig: thermal_wind_maps}
\end{figure*}

\begin{figure*}
   \resizebox{\hsize}{!}
            {\includegraphics[width=\textwidth]{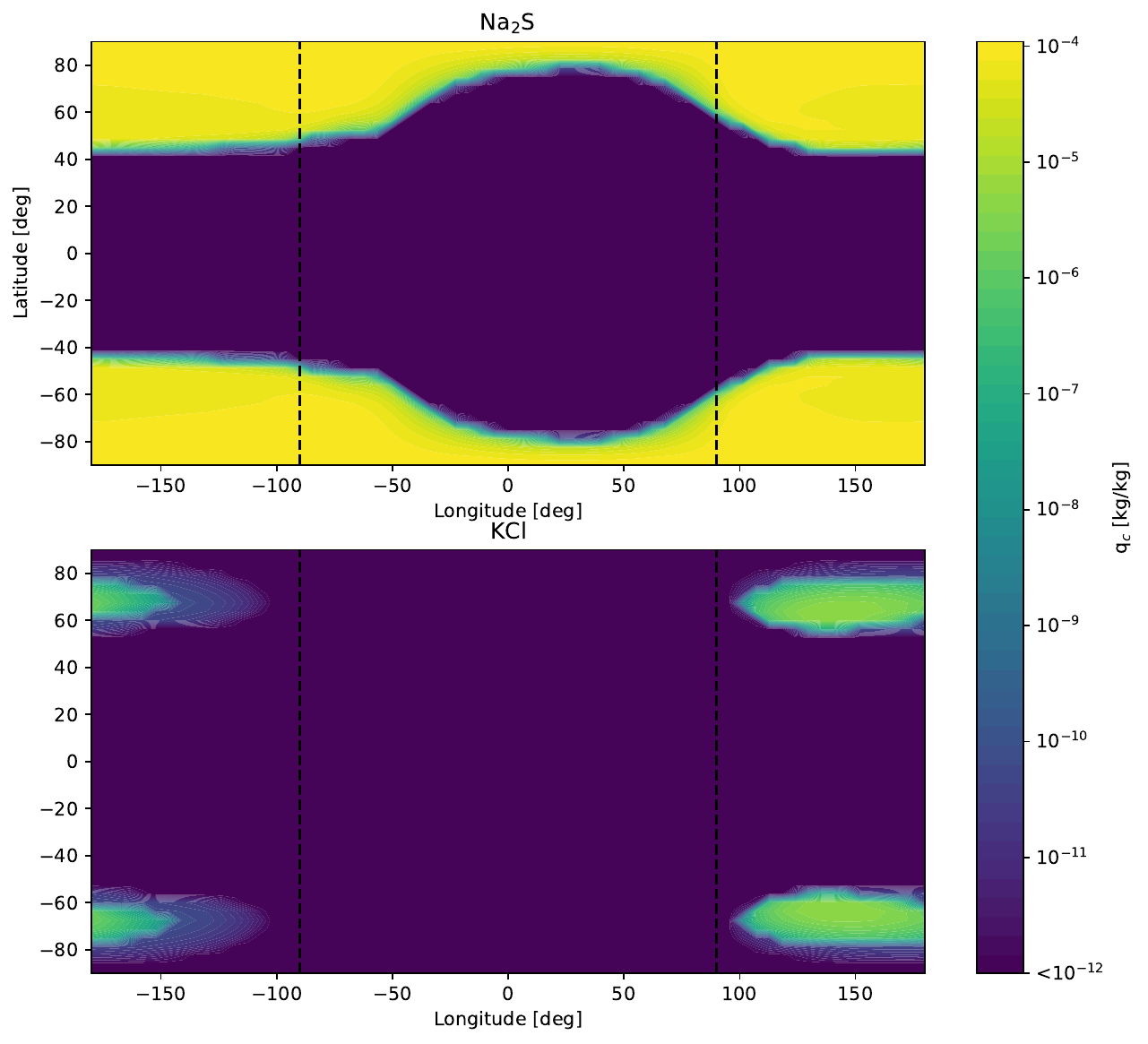}}
      \caption{Horizontal maps of clouds at a pressure of 25 mbars. The top panel displays the \NaS clouds distribution while the bottom panel displays the \kcl distribution. Due to the lower condensation temperature of \kcl, we only observe its formation on the night-side polar vortices whilst \NaS uniformly forms at latitude higher than $\pm$ 40$^{\degree}$ on the nightside, and at higher latitudes on the dayside. }
         \label{fig: clouds_maps}
\end{figure*}

\begin{figure*}
   \resizebox{\hsize}{!}
            {\includegraphics[width=\textwidth]{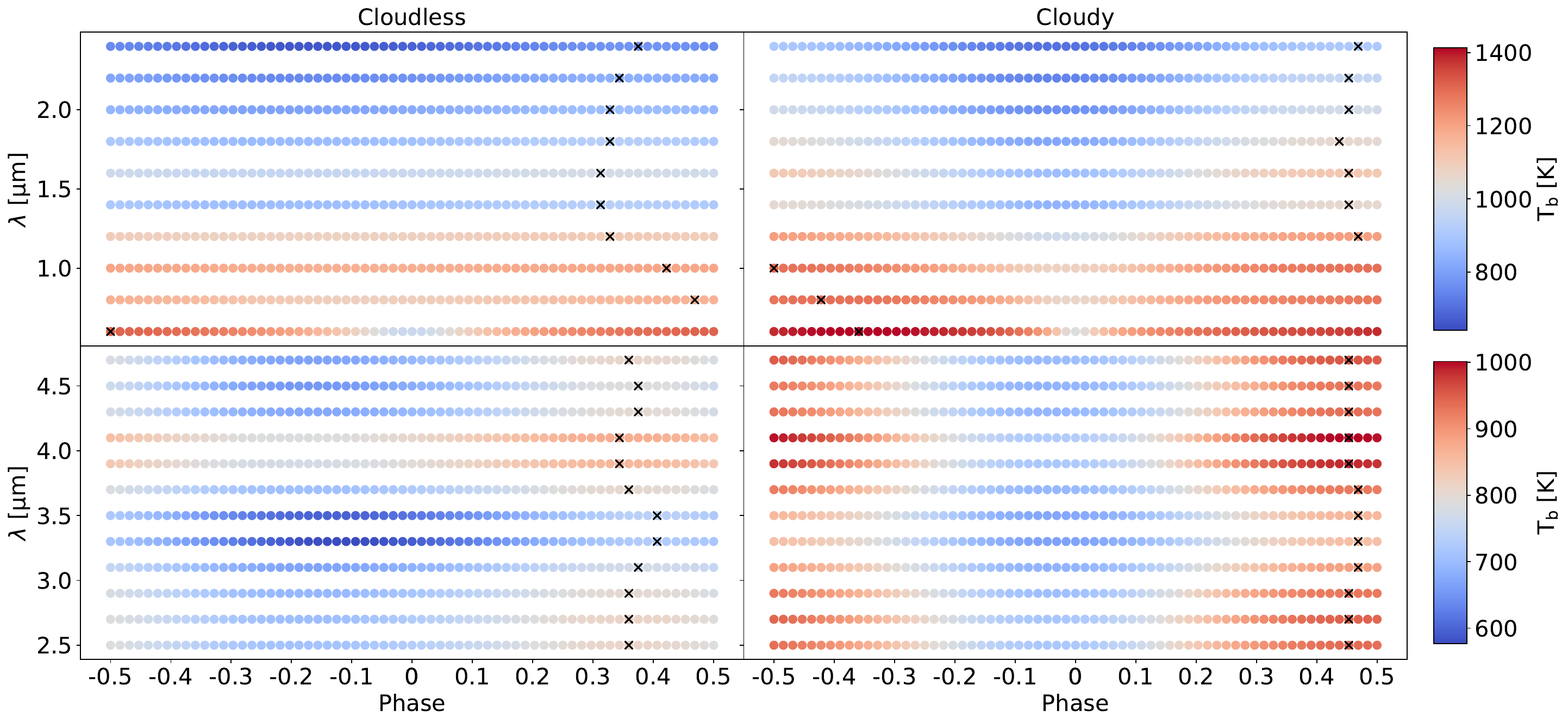}}
      \caption{Cloudless (left) and cloudy (right) phase curves on the NIRSpec-PRISM spectral range, converted to planetary brightness temperature. The black crosses indicate the phase of the maximum flux for a given wavelength (i.e., the spectral offsets). Phase 0 corresponds to the night side (i.e., the transit event).}
         \label{fig: nirspec-phycurve}
\end{figure*}

\begin{figure*}
   \resizebox{\hsize}{!}
            {\includegraphics[width=\textwidth]{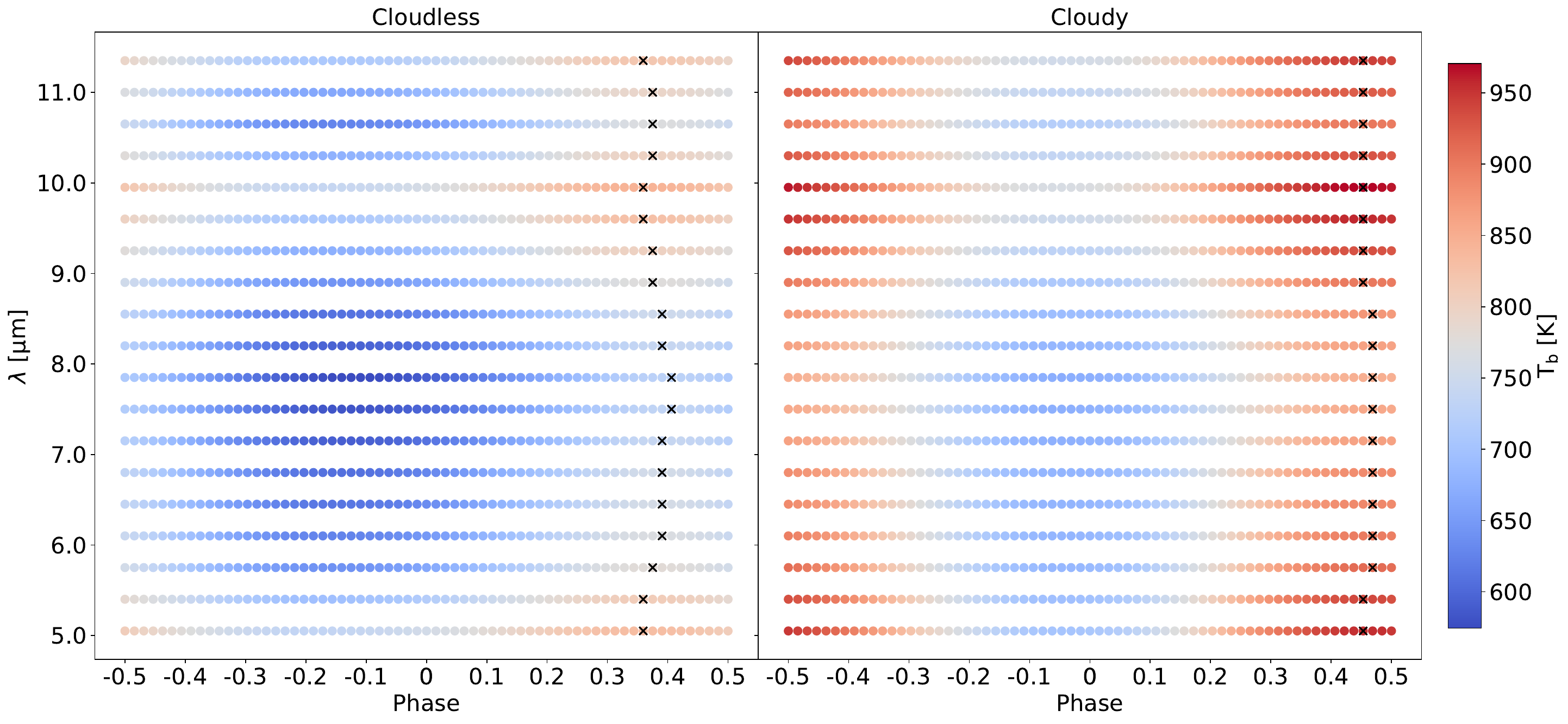}}
      \caption{Cloudless (left) and cloudy (right) phase curves on the MIRI-LRS spectral range, converted to planetary brightness temperature. The black crosses indicate the phase of the maximum flux for a given wavelength (i.e., the spectral offsets). Phase 0 corresponds to the night side (i.e., the transit event).}
         \label{fig: miri-phycurve}
\end{figure*}

In this section, we show a few maps of the dynamical and thermal structure modelled with the \texttt{generic PCM}. Fig. \ref{fig: thermal_wind_maps} shows longitude-latitude temperature maps and zonal-mean zonal winds maps for our cloudless and cloudy simulations. In both cases, we observe a strong day-night temperature contrast and an eastward shift of the hotspot with regard to the sub-stellar location. Between a cloudy and cloudless atmosphere, we observe a reduction in the hotspot offset, correlated with the reduction of the spectral phase curves offset shown in Fig.\ref{fig: nirspec-phycurve},\ref{fig: miri-phycurve}. We also observe a super-rotating, eastward, equatorial jet and a change in the wind structure at high latitudes between the two simulations. This is due to the radiative feedback of clouds, that alter the meridional temperature gradients (see \cite{teinturier_radiative_2024} for a more thorough description). \\ 
We also show longitude-latitude maps of the distribution of the clouds in Fig. \ref{fig: clouds_maps}.

\section{Phase-curves} \label{appendix: phasecurve}   
We show here a subset of the spectral phase curves computed from the cloudy and cloudless simulation on the NIRSpec-PRISM and MIRI-LRS range.
%%%% miri here

\end{appendix}
\end{document}